\documentclass[aps,prl,twocolumn,longbibliography,epsfig,floats,preprintnumbers,floatfix,superscriptaddress]{revtex4-2}
\usepackage{graphicx,amsmath,amssymb,bm,color,sidecap}
\usepackage[latin2]{inputenc}
\usepackage{dcolumn,nicefrac,blindtext,nicefrac,pgfplots}

\begin{document}
\title{Tracking the Morphological Evolution of Neuronal Dendrites by First-Passage Analysis}
\author{Fabian H.\ Kreten}
\affiliation{Department of Theoretical Physics, Saarland University, 66123 Saarbr\"ucken, Germany}
\affiliation{Center for Biophysics, Saarland University, 66123 Saarbr\"ucken, Germany}
\author{Barbara A.\ Niemeyer}
\affiliation{Center for Biophysics, Saarland University, 66123 Saarbr\"ucken, Germany}
\affiliation{Department of Molecular Biophysics, Center for Integrative Physiology and 
Molecular Medicine, School of Medicine, Saarland University, 66421 Homburg, Germany}
\author{Ludger Santen}
\affiliation{Department of Theoretical Physics, Saarland University, 66123 Saarbr\"ucken, Germany}
\affiliation{Center for Biophysics, Saarland University, 66123 Saarbr\"ucken, Germany}
\author{Reza Shaebani}
\email{shaebani@lusi.uni-sb.de}
\affiliation{Department of Theoretical Physics, Saarland University, 66123 Saarbr\"ucken, Germany}
\affiliation{Center for Biophysics, Saarland University, 66123 Saarbr\"ucken, Germany}

\begin{abstract}
A high degree of structural complexity arises in dynamic neuronal 
dendrites due to extensive branching patterns and diverse spine 
morphologies, which enable the nervous system to adjust function, 
construct complex input pathways and thereby enhance the 
computational power of the system. Owing to the determinant role 
of dendrite morphology in the functionality of the nervous system, 
recognition of pathological changes due to neurodegenerative 
disorders is of crucial importance. We show that the statistical 
analysis of a temporary signal generated by cargos that have 
diffusively passed through the complex dendritic structure yields 
vital information about dendrite morphology. As a feasible 
scenario, we propose engineering mRNA-carrying multilamellar 
liposomes to diffusively reach the soma and release mRNAs, which 
are translated into a specific protein upon encountering ribosomes. 
The concentration of this protein over a large population of neurons 
can be externally measured, as a detectable temporary signal. Using 
a stochastic coarse-grained approach for first-passage through 
dendrites, we connect the key morphological properties affected 
by neurodegenerative diseases--- including the density and size 
of spines, the extent of the tree, and the segmental increase 
of dendrite diameter towards soma--- to the characteristics of 
the evolving signal. Thus, we establish a direct link between 
the dendrite morphology and the statistical characteristics of 
the detectable signal. Our approach provides a fast noninvasive 
measurement technique to indirectly extract vital information 
about the morphological evolution of dendrites in the course 
of neurodegenerative disease progression.
\end{abstract}

\maketitle
 
\section*{Introduction}

The elaborate branching morphology of neuronal dendrites in 
advanced nervous systems allows the neuron to interact simultaneously 
with several neighbors through their axon terminals, leading to a 
complex network of signaling pathways \cite{Jan10,Poirazi20}. The 
diverse functions of dendritic trees are reflected in the broad 
variation of their architecture in different neuronal types and 
regions. The complex behavior of higher animals has also been 
attributed to the presence of small membranous protrusions 
called dendritic spines \cite{RasiaFilho23,Yuste04,Berry17,
Hering01}. Functional synapses, as the basic computational 
units of signal transmission, consist of the presynaptic 
release site and dendritic protrusions, harbouring the 
signal recognition and transmission units. Spines play a 
vital role in neural functions such as cognition, memory, 
and learning \cite{Nicoll17,Matsuzaki04,Rogerson14,Bloodgood05,
Hongpaisan07} and often serve as the recipients of excitatory 
and inhibitory inputs in the mammalian brain and undergo dynamic 
structural changes regulated by neuronal activity \cite{Bono17,
Moczulska13}. The morphology of spines plays a crucial role as, 
for example, the shape and size of spine head and neck determine 
the number of postsynaptic receptors and the generated synaptic 
current \cite{Matsuzaki04} and control the electrical and 
biochemical isolation of the spine from the dendrite shaft 
\cite{Araya06,Noguchi05}. 

Aging and several neurodegenerative diseases--- e.g.\ fragile 
X and Down's syndromes, Alzheimer's disease, schizophrenia, and 
autism spectrum disorders--- significantly influence the function 
of the nervous system by altering the morphology of dendrites 
\cite{Sudhof08,Kulkarni12,Luebke10,Penzes11,Dorostkar15}. These 
alterations occur in the overall extent of dendritic trees, the 
population of branches, the thickness and curvature of dendrite 
shafts, and the density, morphology, and spatial distribution of 
spines \cite{Petanjek11,Boros17,Tackenberg09,Giannakopoulos09,
Lewis08,Glausier13,Hutsler10,Toro10,Ford23,Bagni19,Irwin01,Orner14,
Tsai04,Deng16}. On the other hand, reversal of morphological changes 
upon treatment has also been reported \cite{Smith09,Biscaro09}. 
Despite the crucial importance of monitoring the structural 
evolution of dendritic trees and spines to diagnose and predict 
neurodegenerative diseases and to monitor success of potential 
treatments, noninvasive imaging of neuronal dendrites is currently 
infeasible. It is even highly challenging to collect statistically 
adequate structural information from direct invasive imaging due 
to technical limitations: Although image analysis techniques for 
3D reconstruction of dendrites have been improved in recent years 
\cite{BenavidesPiccione13,Turner22}, a high resolution image 
can be achieved by electron microscopy which is a very laborious 
technique and practically inappropriate for spatially large-scale 
investigations. Nevertheless, there exist powerful noninvasive 
techniques which allow real-time tracking of brain activities, 
ranging from electro- and magneto-encephalography for electric 
and magnetic field detection \cite{Nunez06,Hari12} to nuclear 
magnetic resonance spectroscopy, positron emission tomography, 
and magnetic resonance imaging (MRI) for measuring the concentration 
of (neuro-)chemicals \cite{Stanley18,Koolschijn23,Bass23,Liang00}.    

In the absence of a direct efficient method for unraveling the 
microscopic morphology of spines and dendrites, we alternatively 
develop a fast noninvasive technique--- based on processing an 
externally detectable signal generated by a large population of 
neurons--- to indirectly obtain essential structural information. 
A possible realization of such a signal can be the concentration 
of an expressed protein by many neurons in the brain region of 
interest, monitored e.g.\ by the MRI technique. We discuss how 
mRNA-carrying multilamellar liposomes can reach a desired region 
of brain tissue using brain-targeted drug delivery techniques, 
pass through the complex dendrite structures to reach the soma, 
and release their mRNAs which encodes a specific protein upon 
encountering ribosomes (alternatively, liposomes could carry 
siRNA molecules to silence RNA and decrease a specific protein 
concentration). We develop a framework that bridges the diverse 
morphological characteristics of dendrites and the statistical 
properties of the detectable evolving signal (linked to the 
first-passage times of passing through the dendrite structure). 
Our technique enables systematic monitoring of neurodegenerative 
disease or treatment progression for individual patients. 

\section*{Coarse-Grained Dendrite Model} 

\begin{figure}[t]
\begin{center}
\includegraphics[width=0.47\textwidth]{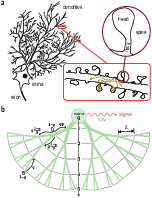}
\caption{Morphology of neuronal dendrites. (a) Schematic drawing 
of a neuron. Insets: (lower) A section of a typical dendritic 
channel. A sample path of a particle is shown; (upper) Structure 
of a mushroom-like spine. (b) Sketch of our binary tree model. 
A tree structure with depth $n\,{=}\,5$ is depicted as an 
example. The possible choices at junctions or dead-ends are 
shown with arrows. The coarse-grained probability to reach a 
neighboring intersect is denoted with $q$, the topological 
bias $p$ represents the segmental increase of dendrite diameter 
towards the soma, and $1{/}t_d$ denotes the pulse emission 
rate.}
\label{Fig:1}
\end{center}
\end{figure}

To model the structure of dendrites, we adopt a mesoscopic perspective and 
consider a regularly branched tree with $n$ generations of junctions. 
Denoting the average distance between adjacent nodes by $L$, the tree 
has a linear extent $n L$. Importantly, as we will discuss, our results 
remain valid under realistic levels of global variation in structural 
parameters or local irregularities across the tree architecture 
\cite{Jose18,Shaebani18}. By coarse-graining the stochastic transport 
within dendritic shafts and spines, we study the dynamics of noninteracting 
particles hopping between the nodes of the tree (see \cite{Jose18} and 
Fig.\,\ref{Fig:1}). Each particle, drawn from an initial reservoir containing 
$\mathcal{N}$ particles, enters the tree at a random node indexed by its 
depth $i\,{\in}[0,n]$, where $i\,{=}\,0$ corresponds to the soma and $i\,{=}
\,n$ to the dead ends. The time of entry is randomly sampled from a geometric 
distribution with mean $t_e$, consistent with the goal of monitoring 
transport behavior over a brain region rather than focusing on individual 
particles. Assuming a uniform probability of arrival across the tree, the 
entry probability increases exponentially with node depth in a binary tree, 
yielding an entry-level distribution $E_{_i}\,{=}\,2^{i{-}1}{/}(2^n{-}1)$ for 
$i{\geq}1$. At each time step, a particle either hops to a neighboring node 
with probability $q$, or remains at its current position with probability 
$1{-}q$. The waiting probability accounts for both stochastic trapping 
inside dendritic spines and the diffusive delay within the dendrite channel. 
We assume that the residence probability inside these biochemical cages 
is depth-independent since the spine number density along the dendrite 
is known to rapidly saturate after a short distance from the soma 
\cite{BenavidesPiccione13,Ballesteros06}. To model the directional preference 
in tracer particle motion toward the soma or dead ends, we introduce a 
topological bias parameter $p$, which governs the directional preference 
of particle movement: transitions toward the soma and toward the dead ends 
occur with probabilities $p$ and $1{-}p$, respectively. Each particle is 
eventually absorbed at the soma. Upon arrival, it emits a transient pulse 
after a randomly sampled delay, referred to as the emission time, which 
is drawn from a geometric distribution with mean $t_d$. While we assume 
absorption at the soma and a specific form of signal generation for 
illustration, the boundary and initial conditions in our framework are 
flexible and can be adapted to match any relevant experimental setup.

Assuming the above dynamics, we can construct a set of coupled master 
equations for the time evolution of the probability $P\!\!_{_i}(t)$ 
for a particle being at depth level $i$ at time $t$, where the hopping 
and trapping probabilities on the network nodes are taken into account; 
see \cite{Kreten25} for details. The resulting dynamics of individual 
particles can be described in general by stochastic two-state models 
\cite{Shaebani19,Shaebani22}. In particular, we previously derived 
the mean first-passage time (MFPT) of being absorbed in the soma (though 
for the initial condition of only entering from the dead ends) in terms 
of the structural parameters $\{n,q,p\}$ by treating the soma as an 
absorbing boundary \cite{Jose18,Shaebani18}. Importantly, we verified 
that the analytical predictions remain valid for realistic degrees of 
structural fluctuations (e.g., diversity in the extent of tree along 
different directions, disorder in the local branching patterns, etc.). 
Although the high sensitivity of the MFPT to the structural characteristics 
of dendrites is promising, MFPT is not a directly measurable quantity 
in dendrites. To realize the practical potential of our approach in 
technology and medicine, here we extend our proposed formalism and 
consider the subsequent steps after the particles reach the soma. We 
assume that each particle emits a temporary pulse after reaching the 
soma. The accumulation of the pulses generated in many neurons results 
in an evolving overall signal intensity $I(t)$ which can be detected 
externally. $I(t)$ contains the information of entering, first-passage, 
and emission times. It indeed reduces to the first-passage time 
distribution (though shifted by two time steps) in the limit of 
instantaneous entering and pulse emission, i.e.\ $t_e\,{=}\,t_d\,{=}
\,1$. For the general case of $t_e{,}\,t_d\,{>}\,1$--- where $I(t)$ 
deviates from the first-passage time distribution--- we demonstrate 
how the statistical characteristics of $I(t)$ can be linked to the 
morphological properties $\{n,q,p\}$ of the dendrite structure. We 
note that the location of signal generation is in principle arbitrary. 
Here, we choose the soma as the signal generation point for simplicity--- 
since this choice reduces the problem to an effective 1D lattice--- 
but the approach is extendable to alternative scenarios with 
different signal generation locations. 

\begin{figure*}[t]
\begin{center}
\includegraphics[width=0.99\textwidth]{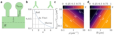}
\caption{Calibration of the mesoscopic model parameters. (a) 
Sketch of a section of the dendrite channel. (b) Bias probability 
$p$ in terms of the allometric exponent $\kappa$. The corresponding 
points for a couple of known structures are marked. The insets show 
schematic drawings of channel diameters at different $\kappa$ regimes. 
(c),(d) Moving probability $q$ in the ($V_\text{head}{+}V_\text{neck}$, 
$\rho$) and ($V_\text{head}{+}V_\text{neck}$, $R$) planes for the 
maximum possible time step $\Delta t\,{=}\,\frac{L^2}{2D}$. Other 
parameters: (c) $R\,{=}\,1 \mu\text{m}$, (d) $\rho\,{=}\,1 \mu
\text{m}^{-1}$. The solid white lines represent isolines of constant 
$q$ with the indicated values. The crosses mark the set of parameter 
values ($V_\text{head}{+}V_\text{neck}\,{=}\,0.55\,\mu\text{m}^3$, 
$\rho\,{=}\,1\,\mu\text{m}^{-1}$, $R\,{=}\,1\,\mu\text{m}$) for a 
typical healthy dendrite as a reference for comparison.}
\label{Fig:2}
\end{center}
\end{figure*}

The details of calibration of the model parameter $q$, clarification 
of the required time resolution of measurements, and estimation of 
the applicability range of our proposed technique are presented in 
the \emph{Supporting Information} file.

\section*{Mapping to Real Dendrite Structures}

We first verify the applicability of our mesoscopic approach 
by mapping the coarse-grained model parameters to the morphological 
characteristics of real dendrite structures. The depth parameter 
$n$ and node-to-node distance $L$ are directly mapped to the extent 
of the dendritic tree, which primarily depends on the nervous system 
and neuronal region and type. For instance, cerebellar Purkinje 
cells of guinea pigs extend up to $200\,\mu\text{m}$ from the soma 
and have ${\sim}\,450$ dendritic terminals \cite{Rapp94}. This 
corresponds to nearly $n\,{=}\,10$ generations of junctions which 
branch out around every $L\,{=}\,20\,\mu\text{m}$.

To map the model parameters $p$ and $q$ to the structure of real 
dendrites, we consider the diffusive dynamics of tracer particles 
along a dendritic tree with protrusions as depicted in Fig.\,\ref{Fig:2}(a). 
The bias $p$ in the direction of motion arises from geometrical 
asymmetries such as tapering of the channel cross-section as well 
as branching at the junctions into $z\,{-}\,1$ children ($z$ being 
the number of branches at each junction). The directional bias can 
be approximated as \cite{Kreten25} 
\begin{equation}
p\,{=}\,\frac{A_p}{A_p\,{+}\,(z\,{-}\,1)\,A_c},
\label{Eq:p1}
\end{equation}
where $A_p$ and $A_c$ denote the cross-sectional areas of the parent 
and child branches, respectively. The areas can be extracted through 
the allometric relation $d_p^{\,\kappa}\,{=}\,\sum_{i{=}1}^{z\,{-}\,
1}d_{c}^{\,\kappa}$ between the diameters $d_p$ and $d_{c}$ of the parent 
and child branches at the junction, where $\kappa$ is the allometric 
exponent \cite{Liao21}. Empirical and theoretical studies suggest 
representative values of the allometric exponent, with $\kappa\,{=}
\,\frac 3 2$ for dendrites of motor neurons [Rall, 1959 \cite{Rall59}], 
$\kappa\,{=}\,2$ for botanical trees [da Vinci's exponent \cite{Grigoriev22}], 
and $\kappa\,{=}\,3$ for vascular and pulmonary networks [Murray's 
exponent\cite{Murray26}]. Other exponents in neuronal context were 
found to be $\kappa\,{=}\,2$ ($p\,{=}\,\frac 1 2$) for Purkinje cells, 
$\kappa\,{=}\,2.5$ ($p\,{=}\,0.47$) for peripheral neuron systems, and 
$\kappa\,{=}\,3$ ($p\,{=}\,0.44$) for axons \cite{DesaiChowdhry22}. Using 
the allometric relation, Eq.\,(\ref{Eq:p1}) results in
\begin{equation}
p\,{=}\,\frac{1}{1+(z\,{-}\,1)^{1{-}\frac{2}{\kappa}}}.
\label{Eq:p2}
\end{equation}
Assuming bifurcations ($z\,{=}\,3$) and symmetric daughter branches yields, 
for example,  estimated values of $p\,{\simeq}\,0.56$, $0.5$ and $0.44$, 
for $\kappa\,{=}\,\frac 3 2$, $2$, and $3$, respectively; see Fig.\,\ref{Fig:2}(b). 
For general channel geometries and driving forces, the bias parameter 
can be obtained by solving a Fick-Jacobs-like equation \cite{Kreten25}. 
Figure\,\ref{Fig:2}(b) illustrates that larger values of $p$ correspond 
to a more pronounced thickening of the channels toward the soma.

To calibrate the probability of motion $q$, we equate the mean time to 
leave one node in the coarse-grained discrete time model with the mean 
travel time from the current junction to any of the neighboring ones 
in the presence of spines. We consider a symmetric branch at which the 
child channels are connected with equal radii and without leaving a void 
space. The entrapment of particles inside spines leads to an effective 
diffusion constant $D\!_{_\text{eff}}\,{=}\,D\,\frac{V_\text{channel}}{
V_\text{channel}{+}V_\text{spines}}$ for the diffusive dynamics inside 
the channel, where $V_\text{channel}$ is the volume of the channel 
segment, $V_\text{spines}$ is the total volume of spines along it, 
and $D$ is the diffusion constant in the absence of spines \cite{Dagdug07}. 
We obtain the following expressions for the moving probability (see 
\emph{Suppl.\,Info.}\ for details):
\begin{equation}
q = \Delta t \displaystyle\frac{2D\!_{_\text{eff}}}{L^2}= \Delta 
t \displaystyle\frac{2D}{L^2}\displaystyle\frac{1}{1 +\displaystyle
\frac{\rho(V_\text{head} +V_\text{neck})}{\pi R^2}},
\label{Eq:q}     
\end{equation}
with $\Delta t$ being the time resolution of measurements, $V_\text{neck}$ 
and $V_\text{head}$ the spine neck and head volumes, $\rho$ the spine 
density per length unit for regularly spaced spines along the channel, 
and $R$ the radius of the channel segment. The above equation imposes 
no explicit bound on $q$, however, both $p$ and $q$ parameters are 
indeed restricted due to the limited biologically relevant ranges of 
the structural parameters. For example, $V_\text{head}{+}V_\text{neck}\,
{\simeq}\,0.5\,\mu\text{m}^3$ and $\rho\,{\simeq}\,1\,\mu\text{m}^{-1}$ 
represent typical structural parameter values for a healthy dendrite 
\cite{Harris92}. We also note that the diffusion constant depends on the 
particle size. Some typical values are: $D\,{\sim}\,20$ (green fluorescent 
protein (GFP) variants inside spines), ${\sim}\,100$ (large $\text{Ca}^{2{+}}$ 
ions inside spines), ${\sim}\,37$ (photo activatable GFP, paGFP, inside 
dendrite channels), and ${\sim}\,23.5\,\mu\text{m}^2{/}\text{s}$ 
(enhanced GFP, eGFP, inside the nucleus) \cite{Bloodgood05,Yasuda11,
Chen02}. Using these values, we obtain $t\,{\simeq}\,0.8$, $0.5$, and 
$0.2\,\text{s}$ for the escape time of eGFP, paGFP, and $\text{Ca}^{2{+}}$ 
from spines and $t\,{\simeq}\,8.5$, $5.4$, and $2.0\,\text{s}$ for their 
travel time from one junction to the next one in a dendritic tree similar 
to that of cerebellar Purkinje cells but in the absence of spines. The 
behavior of $q$ versus the structural parameters of dendrites is shown 
in Figs.\,\ref{Fig:2}(c),(d). It can be seen that $q$ varies monotonically 
with the structural parameters within their biologically relevant ranges, 
which allows for mapping of the morphological characteristics of real 
dendrite structures to the coarse-grained model parameters.

\section*{Signal Processing}

To establish a direct link between the dendrite morphology and the 
statistical characteristics of the detectable signal, our next step 
is to demonstrate how the coarse-grained model parameters influence 
the time evolution of the overall signal. We have recently shown 
that in general the detected signal $I(t)$ from branched structures 
develops a peak followed by an exponential decay at long times 
\cite{Kreten25}. The location and height of the peak and the slope 
of the tail depend on the coarse-grained model parameters and the time 
scales $t_e$ and $t_d$. For small values of $t_e$ and $t_d$, the signal 
intensity is nearly equivalent to the first-passage time distribution 
of passing through the dendritic tree to reach the soma. Note that the 
signal intensity is invariant under the swapping of $t_e$ and $t_d$, 
and the asymptotic behavior is governed by the longest time scale. 
Overall, a faster arrival in the soma and/or a faster emission of 
the signal is associated with an earlier and higher peak and a 
steeper tail of $I(t)$.

To develop a more quantitative understanding of how key model parameters 
govern the signal intensity evolution, we vary $n$, $q$ and $p$ over the 
biologically relevant ranges and calculate various statistical characteristic 
of $I(t)$. Of particular interest is the behavior of the logarithm of the 
median, $\log\!_{_{10}}\!(Q_{\frac12}\!)$. Our previous results revealed 
that the median of the signal intensity varies monotonically in terms of 
the structural parameters, even for large values of $t_e$ and $t_d$ 
\cite{Kreten25}. A similar behavior can be observed for the mean or maximum 
of $I(t)$. Thus, measuring the median of the signal intensity (or any other 
quantity in this category) identifies isosurfaces in the $(n,q,p)$ space 
(i.e.\ a set of admissible $\{n,q,p\}$ values). This is, however, insufficient 
to uniquely determine these parameters. For a unique determination of the 
parameters $\{n,q,p\}$, additional statistical characteristics of $I(t)$ 
whose isolines behave differently from the median ones need to be extracted. 
We tested several quantities, amongst them the variance, skewness, etc. 
We identified a second category of shape quantities which describe the 
dispersion of $I(t)$. The isolines of this category of quantities behave 
differently from the median's ones but not significantly from each other. 
These two independent characteristics of $I(t)$ allow for identifying at 
least a one-dimensional manifold in the $(n,q,p)$ space. As a representative 
of the second category, we choose to work with the relative interquartile 
range $\Delta Q_r$ which is a measure of the statistical dispersion of 
$I(t)$ and has a smooth behavior upon varying the structural parameters. 
It is defined as $\Delta Q_r\,{=}\,\frac{Q_{3{/}4}\,{-}\,Q_{1{/}4}}{Q_{
3{/}4}\,{+}\,Q_{1{/}4}}$, with $Q_{i{/}4}$ being the $i$th quartile (for 
example, the second quartile $Q_{2{/}4}\,{\equiv}\,Q_{1{/}2}$ corresponds 
to the median). We previously verified that the isolines of $\Delta Q_r$ 
and $\log\!_{_{10}}\!(Q_{\frac12}\!)$ have distinctly different orientations 
at small $t_e$ and $t_d$ \cite{Kreten25}. In this regime, each pair of isolines 
for a measured set of $\log\!_{_{10}}\!(Q_{\frac12}\!)$ and $\Delta Q_r$ 
intersect at a single point. However, at longer $t_e$ and $t_d$ time scales, 
the isolines of $\Delta Q_r$ may exhibit a nonmonotonic behavior within 
the relevant range of the structural parameters or become nearly parallel 
to the isolines of $\log\!_{_{10}}\!(Q_{\frac12}\!)$. As a result, each 
pair of isolines for a measured set of $\log\!_{_{10}}\!(Q_{\frac12}\!)$ 
and $\Delta Q_r$ may have several intersections, meaning that $\{n,q,p\}$ 
parameters cannot be uniquely determined. We note that identifying 
the two categories of signal shape parameters may only constrain the 
coarse-grained model parameters to a 1D manifold in the $\{n,q,p\}$ 
space. In general, to uniquely determine the structure, either an 
additional independent quantity can be identified by analyzing 
other shape parameters of $I(t)$ or, alternatively, a constitutive 
relation among the model parameters or the Fourier transform of 
$I(t)$--- known as the empirical characteristic function--- can 
be employed. In the following, we assume for simplicity that the 
two shape parameters, $\log\!_{_{10}}\!(Q_{\frac12}\!)$ and $\Delta Q_r$, 
suffice to uniquely determine the coarse-grained model parameters.

\begin{figure*}[t]
\begin{center}
\includegraphics[width=0.9\textwidth]{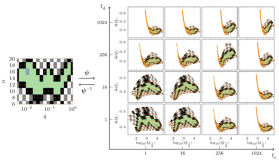}
\caption{Mapping the structural parameters to the characteristics of the 
signal intensity ($\bm\psi$), and vice versa ($\bm\psi^{-1}$). The mapping 
of $(q,n)$ to $\big(\!\log\!_{_{10}}\!(Q_{\frac12}\!),\Delta Q_r\!\big)$ 
(i.e., the logarithm of the median and the relative interquartile range) 
is presented for different mean entering $t_e$ and emission $t_d$ times. 
Other parameter values:\ $p\,{=}\,0.55$. For every marked point on the 
structural parameter domain (yellow crosses), the corresponding location 
on the signal characteristic domain is extracted numerically. The neighborhood 
around each pair of connected points are painted with the same color in 
both domains for clarity.}
\label{Fig:3}
\end{center}
\end{figure*}

\begin{figure*}[t]
\begin{center}
\includegraphics[width=0.9\textwidth]{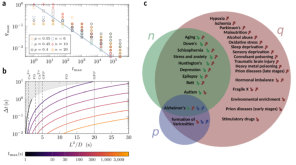}
\caption{(a) Invertibility threshold $q_{_\text{max}}$ versus 
the longest time scale $t_\text{max}{=}\,\text{max}(t_e,t_d)$ 
for different values of $p$ and $n$. The line represents 
$q_{_\text{max}}{=}\,t_\text{max}^{-1{/}2}$. (b) Time 
resolution of measurements $\Delta t$ versus diffusive 
timescale $L^2\,{/}\,D$ for different values of $t_\text{max}$. 
The hatched areas indicate the inadmissible region given by 
$\Delta t{>}L^2{/}2D$, where the probability $q$ would be 
larger than one. The vertical lines mark the relevant range 
along the $x$-axis for Ca\textsuperscript{2+}, fluorescein 
dextran (FD), and green fluorescent protein (GFP) \cite{Yasuda11,
Santamaria06}. (c) Pathologies of dendrite morphology, presented 
in terms of the mesoscopic model parameters $n$, $p$, and $q$ 
(green, blue, and red colors, respectively). Each arrow indicates 
the increase or decrease of the corresponding parameter in the 
course of progression of the given disease. See \cite{Kulkarni12,
Fiala02,Maiti15} and references in the main text.}
\label{Fig:4}
\end{center}
\end{figure*}

The sensitivity of the relation between the two sets of structural 
and signal intensity parameters to the choice of $t_e$ and $t_d$ 
time scales can be more clearly presented in terms of the degree 
of information compression when mapping the phase spaces of these 
two sets to each other. We denote the mapping of structure to 
signal and vice versa with $\bm\psi$ and $\bm\psi^{-1}$, 
respectively. Thus, the connection between the two sets can be 
represented as $\{\log\!_{_{10}}\!(Q_{\frac12}\!), \Delta Q_r
\!\}\,{=}\,\bm\psi(n,q,p)$ and $\{n,q,p\}\,{=}\,\bm\psi^{-1}(\log
\!_{_{10}}\!(Q_{\frac12}\!), \Delta Q_r)$ in general. As an example, 
in Fig.\,\ref{Fig:3} we show the mapping of the $(q,n)$ plane to 
the $\big(\!\log\!_{_{10}}\!(Q_{\frac12}\!), \Delta Q_r\!\big)$ 
plane for different mean entering $t_e$ and emission $t_d$ times. 
We regularly sample the phase space of structural parameters (yellow 
crosses in left panels) and perform extensive simulations to obtain 
the signal intensity $I(t)$ and extract its median $\log\!_{_{10}}\!
(Q_{\frac12}\!)$ and dispersion $\Delta Q_r$ for each sampled set 
of $\{n,q,p\}$. For small values of $t_e$ and $t_d$, the structure 
domain is mapped one-to-one to the signal domain. The mapping 
consists of a slightly skewed rotation but mapping of two or more 
distinct points of the $\{n,q,p\}$ parameter space to a same point 
in the $\{\log\!_{_{10}}\!(Q_{ \frac12}\!), \Delta Q_r\}$ space is 
unlikely, i.e., the map can be inverted. With increasing $t_e$ and 
$t_d$, the high $q$ regions in the structure domain gradually map 
to highly narrow regions in the signal domain. The compression of 
information to a space with one less dimension in the limit of large 
$t_e$ and $t_d$ time scales means that the map cannot be fully inverted 
anymore. To assess the invertibility limit of $q$, we quantify the 
compression of the points by the mapping from structure to signal 
domain; see \emph{Suppl.\,Info.}\ for details. By setting a threshold 
level for the information compression, we can determine the maximum 
value of $q$ (denoted by $q_\text{max}$) up to which the map remains 
invertible (Suppl.\,Fig.\,S1). In Fig.\,\ref{Fig:4}(a), $q_\text{max}$ 
is plotted as a function of $t_\text{max}\,{=}\,\max(t_e,t_d)$, 
i.e., the longest time scale among the entering and emission 
times. It reveals that $q_\text{max}$ decays with $t_\text{max}$ 
roughly as a power-law, which can be presented as $q_\text{max}
\,{\simeq}\,\sqrt{\frac{\Delta t}{t_\text{max}}}$ using the fact 
that $t_\text{max}$ is measured in units of $\Delta t$. On the 
other hand, from Eq.\,[\ref{Eq:q}] the maximum value of $q$ for 
a given dendritic tree is obtained if spines are absent, leading 
to $q_\text{spineless} = \Delta t \frac{2D}{L^2}$ with $D$ being 
the diffusion coefficient in the smooth channel without spines. 
The full range of $q$ is invertible if $q_\text{spineless}\,{\leq}
\,q_\text{max}$, which imposes the constraint
\begin{equation}
\Delta t \leq \left(\frac{L^2}{2D}\right)^2 \frac{1}{t_\text{max}}
\label{Eq:dtThreshold}
\end{equation}
on $\Delta t$, as plotted in Fig.\,\ref{Fig:4}(b) for different values 
of $t_\text{max}$. For a given set of dendritic tree structure and 
tracer particle, the required time resolution of measurements $\Delta t$ 
is inversely proportional to $t_\text{max}$. The vertical lines in 
Fig.\,\ref{Fig:4}(b) mark the relevant range of $\frac{L^2}{D}$ 
for realistic values of the branching distance $L$ and diffusion 
coefficients $D$ for Ca\textsuperscript{2+}, fluorescein dextran 
(FD), and green fluorescent protein (GFP), as a few examples. It 
shows that a slower diffusion of tracer particles and shorter entering 
and emission times lead to a broader possible range for the time 
resolution of measurements $\Delta t$. For time resolutions around 
$\Delta t\,{\simeq}\,0.04\,\text{seconds}$ (typical for currently available 
cameras) and tracer particles with diffusion coefficients similar to GFP, the 
required entering and emission times can be up to a few minutes. However, in 
special techniques such as nuclear magnetic resonance spectroscopy, one deals 
with time resolutions of several seconds, demanding more slowly diffusing 
particles and shorter entering and emission times on a sub-minute scale.

\section*{Morphological Changes during Disease Progression}

The morphology of dendrites is broadly affected by aging 
\cite{Petanjek11,Orner14,Penzes11} or neurodegenerative disorders 
such as Alzheimer's disease \cite{Boros17,Tackenberg09,Giannakopoulos09,
Penzes11,Dorostkar15}, autism spectrum disorders \cite{Hutsler10,Toro10,Ford23,
Bagni19,Penzes11}, epilepsy disorder \cite{CaznokSilveira24}, schizophrenia 
\cite{Lewis08,Glausier13,Penzes11}, Down's syndrome \cite{MarinPadilla72}, 
fragile X syndrome \cite{Irwin01,Bagni19}, prion diseases \cite{Fuhrmann07}, 
and stress-related disorders \cite{Christoffel11}. The affected morphological 
properties include the overall extent of dendritic tree, the segmental 
increase of dendrite diameter towards the soma, the population of branches, 
the thickness and curvature of dendrite shafts, and the morphology, 
density, and spatial distribution of spines. Here we clarify how these 
morphological changes influence our mesoscopic model parameters $\{n,q,p\}$. 
This information is then used in Fig.\,\ref{Fig:4}(c) to categorize 
the neurodegenerative disorders--- those for which clear trends for 
the pathological changes of dendrite structure have been reported in 
the literature--- based on the expected trends of the model parameters 
in the course of disease progression. 

Variation of the extent of dendritic tree trivially influences the depth 
parameter $n$. Reduction of the tree extent is the observed trend during 
aging and several disorders such as Down's syndrome, schizophrenia, 
Alzheimer's disease, autism spectrum disorders, epilepsy disorder, 
stress-related disorders, Huntington's disease, etc. Increasing 
of the tree extent due to neurodegenerative disorders has not 
been reported to our knowledge.

The moving probability $q$, given by Eq.\,(\ref{Eq:q}), is the only 
parameter affected by the presence of spines. $q$ increases with 
decreasing spine size or density as observed, e.g., in aging, Down's 
syndrome, Alzheimer's disease, and schizophrenia; see Fig.\,\ref{Fig:4}(c) 
for an extended list of relevant disorders. Conversely, the spine density 
increases in a few cases such as autism spectrum disorders, fragile X 
syndrome, and hormonal imbalance, leading to the decrease of $q$. 
Nevertheless, the pathology of fragile X makes the prediction of 
$q$ variations complicated: The increase of spine density (decrease 
of $q$) is accompanied by the shrinkage of spine size (increase of 
$q$); thus, the two effects compete and may even compensate each 
other such that $q$ remains unchanged. The influence of hormonal 
imbalance on $q$ depends on the hormone type and whether there 
is a deficiency or surplus.

The disorders that change the width of dendrite shafts influence 
the bias parameter $p$ in general. Particularly, the decrease 
in dendritic arborization is often correlated with the overall 
reduction of the channel width. The details of width reduction 
determine the direction of changes of $p$: While a uniform 
reduction of channel radii effectively increases $p$ (and 
decreases $q$), a radius-dependent reduction may change $p$ 
in both directions. Moreover, an inhomogeneous spatial pattern 
of $p$ can be caused by local changes of the channel width. 
For instance, local thinning of channel occurs in the vicinity 
of amyloid plaque deposition in Alzheimer's disease. Finally, 
the disorders that reduce the population of branches can 
increase the average node-to-node distance $L$, resulting 
in smaller $q$ and $n$. 

We note that the pathology of spine and dendrite structure is more 
complicated in other diseases. For instance, distortion of spine 
shape observed in most mental retardations makes the prediction 
of the trend of $q$ difficult. Despite the currently available 
information discussed above, there is a lack of quantitative 
studies to clarify the impact of various neurodegenerative 
disorders on dendritic spine, tree metric, and topological 
morphology.

\begin{figure*}[t]
\begin{center}
\includegraphics[width=0.99\textwidth]{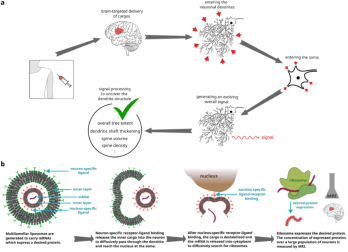}
\caption{(a) Schematic of our proposed noninvasive technique for a 
fast indirect measurement of the structural properties of dendrites 
based on processing a signal generated by a large population of 
neurons. By conducting regular measurements for a given patient, 
essential information about the morphological evolution of dendrites 
in the course of neurodegenerative disease or treatment progression 
can be extracted. (b) Schematic illustration of multilamellar liposomes 
designed to generate a specific protein. As a detectable temporary 
signal, the concentration of the expressed protein across the brain 
region of interest is externally monitored by the MRI technique.}
\label{Fig:5}
\end{center}
\end{figure*}

\section*{Discussion}

We have proven that the parameters $\{n,q,p\}$ of our mesoscopic 
model can be extracted by analyzing the detectable temporarily signal 
generated by a large population of neurons, provided that the time scales 
of entering the dendrites and emission of signal after reaching the soma 
are sufficiently small compared to the mean first-passage time of passing 
the dendritic tree. Although we constrained our analysis to signals formed 
by spontaneous pulses emitted by the particles in the soma with activation 
probability $\frac{1}{t_d}$, signals of other forms can be easily obtained 
from the signal studied in the present work. For example, if one seeks 
insight into the ability of neurons to integrate spine-derived (concentration) 
signals, the number of particles in the soma that have not yet emitted 
their pulse (i.e.\ are still active in this case) would be of interest.
This quantity at time $t$ is given by $t_d\,I(t{+}1)$, i.e.\ our measured 
signal shifted by one time step to the left and scaled by the mean 
activation time. In Suppl.\,Fig.\,S3, we present this signal alongside 
the time evolution of the fraction of particles in the soma that have 
not yet emitted their pulse for different values of $t_e$ and $t_d$ and 
for healthy versus differently degenerated dendritic trees.

On the other hand, the model parameters $\{n,q,p\}$ can be directly 
linked to the morphology of real dendrites via Eqs.\,(\ref{Eq:p2}) and 
(\ref{Eq:q}). Since there are several morphological characteristics 
on the right hand sides of these equations, they cannot be uniquely 
determined by a given set of $\{n,q,p\}$. Nevertheless, most 
neurodegenerative disorders affect only a few of the morphological 
properties of dendrites. Therefore, by conducting regular patient 
monitoring for a given disease, the observed changes in the parameters 
$\{n,q,p\}$ can be attributed to the changes of the morphological 
properties relevant to that specific disease. For instance, the growth 
rate of $q$ and reduction rate of $n$ for a patient with schizophrenia 
reflect, respectively, how fast the mean spine volume and the extent 
of dendritic tree are shrinking over time. 

To link the detected signal intensity to the mesoscopic model parameters 
we have considered an ideal regular tree structure, while real dendritic 
trees are irregularly branched, spines have diverse sizes, and their 
spatial distribution is inhomogeneous. These fluctuations naturally 
cause variations in the corresponding model parameters $\{n,q,p\}$. 
However, we verified in our previous study \cite{Jose18} that the 
analytical results for the first-passage times of passing a regular 
tree structure remain valid when realistic degrees of global fluctuations 
of the structural parameters across the tree or local structural 
irregularities in the branching patterns are considered. Since 
the dependence of the signal intensity on the dendrite morphology 
is due to the contribution of the fist-passage times (and not the 
entering $t_e$ and emission $t_d$ times), we conclude that the 
presented results in the current study remain valid under typical 
structural irregularities and fluctuations observed in real neuronal 
dendrites.

We have characterized the behavior of the signal intensity $I(t)$ 
by two quantities, the logarithm of the median $\log\!_{_{10}}\!(
Q_{\frac12}\!)$ and the relative interquartile range $\Delta Q_r$. 
The former is a representative of a category of the statistical 
measures including the mean, median, and maximum of $I(t)$. The 
latter quantifies the statistical dispersion of $I(t)$ and behaves 
similar to quantities such as the normalized variance and skewness. 
One may still identify further independent quantities by analyzing 
other moments of $I(t)$. Additional statistical measures of $I(t)$--- 
which vary smoothly with the parameters $\{n,q,p\}$ and exhibit 
isolines that differ from those of $\log\!_{_{10}}\!(Q_{\frac12}\!)$ 
and $\Delta Q_r$--- can in principle improve the accuracy of the 
extracted values of the model parameters $\{n,q,p\}$.

Our study focuses on solving the technical problem of processing the 
generated signal and linking it to the dendrite morphology. Nevertheless, 
preliminary steps should first be taken to generate the signal. This 
includes the transport of cargos to desired regions of brain tissue, 
entering the dendrites, travelling to the soma, and generating a temporary 
signal. Our proposed measurement procedure is summarized in Fig.\,\ref{Fig:5}(a). 
While addressing the technical issues of each preliminary step is 
beyond the scope of this study, obtaining a detectable signal is 
indeed feasible using the currently available technologies. In the 
following, we discuss possible methods for each of the measurement 
procedure steps:
\begin{enumerate}
\item {\bf Transport of cargos to desired regions of brain tissue by 
means of brain-targeted drug delivery techniques.} Promising strategies 
have been developed so far to deliver drugs specifically to the brain 
to treat neurological disorders while minimizing systemic side effects 
\cite{Saraiva16,JuilleratJeanneret08,Zhang19}. Some of the currently 
feasible techniques include nanoparticle-based delivery \cite{Yan24,
Saraiva16,Nance12,Blasi07,Gajbhiye20,Roney05,Schnyder05,Thomsen15,
Spuch11}, focused ultrasound (noninvasive technique which offers 
spatially targeted drug delivery by transiently disrupting the 
blood-brain barrier (BBB) to pass through) \cite{Hynynen05}, and 
carrier-mediated transport (utilizing endogenous transport systems 
like glucose or amino acid transporters facilitates drug transport 
across the BBB. Drug molecules are conjugated with ligands that 
target these transporters to enhance brain uptake) \cite{Pardridge05}.
Moreover, injection into the spinal chord fluid or into ventricles 
could be an option to pass the BBB as well. 
\item {\bf Entering the neuronal dendrites and passing through their 
complex structure to reach the soma.} After reaching the area around 
the neurons, the contents of the cargos can enter the neuron by means 
of neuron-specific receptor-ligand binding \cite{Blasi07,Gajbhiye20,
Roney05,Schnyder05,Thomsen15,Spuch11}. This would mainly occur through 
the dendritic tree rather than axon or soma since the outer area of 
the neuron is mainly formed by the dendritic tree. Nevertheless, the 
contribution of entering from soma or axon to the generated signal 
can be evaluated and subtracted, as long as the axons and somata 
do not undergo morphological changes in the course of disease 
progression or treatment.

\item {\bf Generating a temporary signal and detecting it.} Here,
we mean any kind of detectable signal such as, but not limited to, 
electric or magnetic fields generated by many neurons. There are 
powerful noninvasive techniques for real-time tracking of brain 
activities. Electro- and magneto-encephalography for electric 
and magnetic field detection are established neurotherapeutic 
tools \cite{Nunez06,Hari12}. Another possibility is to employ 
nuclear magnetic resonance spectroscopy (NMRS), which allows 
for noninvasive measurements of the concentration of different 
neuro-chemicals within a volume of brain down to a few cubic 
centimeters \cite{Stanley18,Koolschijn23}. The concentrations 
of substances generated in the somata of neurons can be obtained 
via NMRS with a time resolution of a few seconds which, depending 
on the diffusion constant of the particles, can remain within 
the feasibility range of our proposed method \cite{Gussew10}. 
Positron emission tomography and magnetic resonance imaging 
(MRI) can be also employed to measure the concentration of 
neuro-chemicals \cite{Bass23,Liang00}. 

\item {\bf Processing the evolving overall signal.} After detecting 
the signal, the approach developed in this paper enables one to 
process the signal intensity to uncover the morphology of dendrites.
\end{enumerate}

As a more detailed plan for generating a detectable signal, we 
propose a protein expression scenario by injecting specific mRNAs 
carried by multilamellar liposomes; see Fig.\,\ref{Fig:5}(b). 
The concept of producing multilamellar liposomes is currently 
feasible and has been realized in the context of cell activity 
regulation, immunotherapy, and vaccination \cite{Tenchov21,Shi23,
Moon11}. Transport of liposomes to desired regions of brain and 
uptake of them by neurons have been feasible by modifying 
their surface with ligands targeting specific receptors on 
brain endothelial cells or neuronal dendrites \cite{Blasi07,
Gajbhiye20,Roney05,Schnyder05,Thomsen15,Spuch11}. Upon 
neuron-specific receptor-ligand binding, the multilamellar 
liposome enters the dendrite and loses its outer layer, leading 
to the release of the inner cargo into the cytoplasm. To 
enhance the dendrite specific entering, the endocytosis 
events around the synapses can be harnessed. For example, 
there is evidence that AMPA receptors are preferentially 
endocytosed around synapses \cite{Rosendale17}. The inner cargo 
is conjugated with nucleus-specific ligands \cite{Jang21,
Sharma22,DSouza23} and diffuses inside the dendritic tree until 
it enters the soma and reaches the nucleus. The cargo can be 
designed to be destabilized or dissolved after the nucleus-specific 
receptor-ligand binding pins it to the exterior of the nucleus. 
This can be achieved, e.g., through specific proteolytic enzymes 
or pH-sensitive components in the cargo structure or adjusting 
the concentration of aqueous ionic solutions inside the cargo 
\cite{Chu23} to respond to the environmental differences between 
the region around the nucleus and the rest of the cytoplasm. The 
destabilized cargo releases mRNAs into the cytoplasm, which will 
diffusively search for ribosomes to produce a desired protein, 
such as ferritin. A typical neuron contains millions of ribosomes 
but their homeostatic distribution is still unknown \cite{Dastidar22,
Fusco21} (though recent studies revealed spacial inhomogeneities 
in the protein translation across the neuron, attributed to the 
spacial distribution of mRNAs and potential local specialization 
of ribosomes \cite{Fusco21,Glock21}). The expressed protein 
should be harmless and degrade over a reasonable time. Variations 
of the concentration of this protein over a large population of 
cells can be externally detected. For example, expression of 
ferritin can be monitored by the MRI technique \cite{Liang00}. 
We note that the presented formalism in the previous sections 
to obtain the first-passage time distribution of reaching the 
soma can be straightforwardly extended to calculate the additional 
first-passage time distribution of reaching from the soma to the 
distributed ribosomes. Moreover, here we considered a spontaneous 
signal emission but the formalism can be adapted to other scenarios 
such as a gradually degrading signal.   

The clinical translation of these proposed techniques certainly 
requires rigorous testing and validation to ensure the safety 
and specificity of the novel approaches. Uptake and transport 
of cargos can be initially tested, e.g., in cultured murine neurons. 
For the plan proposed in Fig.\,\ref{Fig:5}(b) based on the expression 
of proteins by ribosomes, spatial distribution of ribosomes in 
different types of healthy neurons needs to be determined. We 
expect that our proposals can potentially trigger active research 
and development in the fields of neuroscience, molecular engineering, 
and pharmacology. 

To conclude, a framework has been developed to link the statistical 
characteristics of a detectable signal generated after reaching the 
somata of neurons to the morphological properties of neuronal dendrite 
structures. Our results open the possibility of indirectly 
monitoring the morphological evolution of dendrites in the 
course of neurodegenerative disorder progression. The mesoscopic 
approach presented in this study can be generalized to cope 
with further details of transport in real dendrite structures, 
such as handling the memory effects and aging inside spines 
\cite{Sadjadi21} or to include active transport of cargos 
along microtubules \cite{Hafner16}. Besides drawing conclusions 
regarding the morphological changes of dendrite structures, 
investigation of the first-passage properties of stochastic 
motion inside dendrites can deliver vital information about 
the ability to preserve local concentrations or induce 
concentration gradients of ions and molecules. These are tightly 
connected to neural functions and allow for drawing important 
physiological conclusions. The proposed approach also provides 
a route into a variety of other stochastic transport phenomena 
e.g.\ in varying energy landscapes, branched macromolecules and 
polymers, and labyrinthine environments with absorbing boundaries.

\bibliography{Refs-Dendrites}

\newpage

\begin{widetext}

\vspace{20mm}

\begin{center}
{\LARGE Supplementary Information to Tracking the Morphological Evolution of Neuronal 
Dendrites by First-Passage Analysis}\\
\end{center}

\subsection*{Mapping dendritic structure to the model parameter $q$}

The mean escape time $\langle t \rangle$ from a junction to any neighboring furcation 
can be expressed, on the one hand, in terms of channel geometry as $\langle t \rangle
\,{=}\,\frac{L^2}{2 D}$, assuming diffusive dynamics. On the other hand, using the 
discrete-time framework of the model with observation time resolution $\Delta t$, 
it is given by $\langle t \rangle\,{=}\,\frac{\Delta t}{q}$. Equating these two 
expressions yields a relation between the moving probability $q$ and the geometric 
parameters of the dendritic structure, $q\,{=}\,\Delta t\,\frac{2D}{L^2}$. However, 
we have assumed a smooth channel so far, thus, this relation does not yet account 
for the effects of trapping in dendritic spines. In the following, we show how 
these effects can be incorporated into the framework.
\smallskip\smallskip\smallskip

We note that transient trapping events along the channel do not induce any bias 
in the motion towards one end of the channel segment, thus, no modification is 
required in the calibration relation for the parameter $p$. However, for the 
escape time $\langle t \rangle$, frequent interruption of motion by entrapment 
events in spines has a considerable impact. To keep the model traceable, this 
impact is taken into account by an effective asymptotic diffusion constant 
$D\!_{_\text{eff}}$. Previous studies have already calculated such an effective 
diffusion constant in a geometry almost tailored to the diffusive transport 
in spiny dendrites \cite{Dagdug07}. The geometry considered there consists of 
a cylindrical tube from which identical spines protrude periodically. The spines 
were modeled as spherical cavities connected to the main shaft by narrow cylindrical 
necks. The effective diffusion constant derived in \cite{Dagdug07}, adapted to 
our application, is given by
\begin{equation}
D\!_{_\text{eff}} = D\frac{V_\text{channel}}{V_\text{channel} + 
V_\text{spines}} = D\frac{1}{1 +\frac{V_\text{spines}}{V_\text{channel}}},
\end{equation}
where $D$ is the diffusion constant without protrusions, $V_\text{channel}$ 
the channel volume, and $V_\text{spines}$ the total volume of spines. Let 
us assume a uniform distribution of spines along the dendritic channel with 
the density $\rho$ per length unit. Denoting the spine head volume with 
$V_\text{head}$ and neck volume with $V_\text{neck}$, the ratio 
$\frac{V_\text{spines}}{V_\text{channel}}$ can be written as
\begin{equation}
\frac{V_\text{spines}}{V_\text{channel}} = \frac{\rho(V_\text{head} + 
V_\text{neck})}{\pi R^2},
\end{equation}
where $R$ is the channel radius. Substituting $D\!_{_\text{eff}}$ into 
the calibration relation for $q$ yields
\begin{align}
q = \Delta t \frac{2D}{L^2}\frac{V_\text{channel}}{V_\text{channel}+
V_\text{spines}}= \Delta t \frac{2D}{L^2}\frac{1}{1 +\frac{\rho
(V_\text{head} +V_\text{neck})}{\pi R^2}}.  
\label{Eq:q1}     
\end{align}
As $q$ is a probability, it cannot be larger than one, imposing a constraint 
on the time resolution of observation $\Delta t$. Since the relation $\frac{
V_\text{channel}}{V_\text{channel} +V_\text{spines}} \leq 1$ always holds, 
the condition
\begin{equation}
\Delta t \leq \frac{L^2}{2D}
\end{equation}
ensures that $q$ always remains as a valid probability, i.e., 
$q\,{\leq}\,1$.
\smallskip\smallskip\smallskip

To derive the above calibration relation, we have made a few simplifying 
assumptions for the diffusive dynamics of tracer particles inside dendritic 
channels. For example, a constant distance $L$ between successive junctions 
is assumed. However, $L$ may vary in real dendrite structures, not only 
between the segments within one generation but also between different 
generations. The primary segments of apical dendrites of pyramidal neurons 
and terminal segments of all dendrites are reported to be longer than 
intermediate segments. As a result, $q$ should slightly vary with the 
depth of dendritic tree due to its $L$-dependence. 
\smallskip\smallskip\smallskip

Additionally, spines are inhomogeneously distributed over the dendritic 
tree. There are almost no spines very close to the soma but the spine 
number density rapidly grows and saturates after a short distance from 
the soma. However, the gradual thinning of the channel towards dendritic 
terminals practically increases the trapping probability inside spines 
and, hence, decreases $q$. 
\smallskip\smallskip\smallskip

Nevertheless, we verified in our previous study \cite{Jose18} that the 
analytical results for the first-passage times of passing a regular 
tree structure remain valid when realistic degrees of global fluctuations 
of the structural parameters across the tree or local structural 
irregularities in the local branching patterns are considered. To 
conclude, the structural parameters which enter into the calibration 
relations for $q$ and $p$ parameters should be considered as average 
values over the entire dendritic tree.

\section*{Considerations for choosing the time resolution of measurements}

In this section we discuss the choice of the time resolution of measurements 
$\Delta t$ and provide a rough estimate of the applicability range of our 
proposed method. Figure\,3 of the manuscript revealed that the invertibility 
of mapping the structural parameters to the signal characteristics breaks 
down for large entering $t_e$ and/or emission $t_d$ times. We further 
clarified that only the mapping of high $q$ regions is problematic, while 
low $q$ regions can be resolved even for very large values of $t_e$ and 
$t_d$. To assess the invertibility limit of $q$, in the following we 
quantify the compression of the points by the mapping. For any given 
point in the structural parameter space and the corresponding point 
in the signal domain, the degree of compression is determined by 
calculating two distances: the minimum distance $\ell_{\text{struct}}$ 
between the selected point and all other points in the structural 
parameter space and the minimum distance $\ell_{\text{signal}}$ 
between the corresponding point and all other points in the $\big(\!
\log\!_{_{10}}\!(Q_{\frac12}\!), \Delta Q_r\!\big)$ plane of signal 
characteristics. In each of the two domains, the distance between 
a pair of points $(x_1,y_1)$ and $(x_2,y_2)$ is calculated using 
the metric
\begin{equation}
\delta\big(\!\left(\begin{smallmatrix} x_1 \\ y_1 \end{smallmatrix}\right), 
\left(\begin{smallmatrix} x_2 \\ y_2 \end{smallmatrix}\right)\!\big)
= \sqrt{\left(\frac{x_2 -x_1}{u_x}\right)^2 +\left(\frac{y_2 -y_1}{u_y}\right)^2},
\label{Eq:ScaledMetric}
\end{equation}
where $(x_i,y_i)$ can be any pair of the structural parameters $\{p, q, n\}$ 
or a point in the $\big(\!\log\!_{_{10}}\!(Q_{\frac12}\!), \Delta Q_r\!\big)$ 
plane of signal characteristics. $u_x$ and $u_y$ denote the total variation 
range along $x$ and $y$ axes, respectively. Note that $u_x$ and $u_y$ in 
the signal domain are determined by combining the variation range of $\log
\!_{_{10}}\!(Q_{\frac12}\!)$ or $\Delta Q_r$ over all choices of the entering 
and emission times.
\smallskip\smallskip\smallskip

From the minimum distances $\ell_{\text{struct}}$ and $\ell_{\text{signal}}$, 
the volumes of the neighbourhoods in the two domains can be estimated 
as $\ell_{\text{struct}}^2$ and $\ell_{\text{signal}}^2$, respectively. 
We introduce the ratio of the volumes $C\,{=}\,\frac{\ell_{\text{signal}
}^2}{\ell_{\text{struct}}^2}$ (hereafter refereed to as compression ratio) 
as a measure of the degree of mapping compression. An invertible mapping 
requires $C\,{>}\,0$. In this regard, the compression ratio has similarities 
with the Jacobian determinant of the mapping. The difference between them 
is that the distances to all other points in each domain are taken into 
account in $C$ whereas for the Jacobian only the neighborhood in each 
domain enters. Therefore, $C$ is a stronger measure for invertibility 
because it vanishes even when a point far away from the one where $C$ 
is calculated is mapped to the same point in the signal domain, but the 
Jacobian determinant cannot capture it due to its local nature. In the 
case of a globally invertible map, $C$ is an estimate for the absolute 
value of the Jacobian determinant.
\smallskip\smallskip\smallskip

Setting a lower compression threshold $C_\text{min}$ allows us to identify 
the points for which the invertibility of the map between structural and signal 
domains practically breaks down. By choosing a threshold value $C_\text{min}
\,{=}\,0.05$, we identify the points where mapping the structural parameters 
to the signal characteristics plane is not invertible. This procedure is 
visualized in Fig.\,S1 for a constant $p$ and various entering and emission 
times. Next, we determine $q_\text{max}$ as the maximum value of $q$ up to 
which the map is invertible for all values of the other dimension of structural 
parameters ($n$ in the cases presented in Fig.\,S1). We checked that the choice 
of the threshold level $C_\text{min}$ has no qualitative impact on the behavior 
of $q_\text{max}$ and only induces minor quantitative changes.
\smallskip\smallskip\smallskip

In Fig.\,S2, $q_\text{max}$ is plotted as a function of the entering time 
$t_e$ and the emission time $t_d$ for a given value of $p$. It can be seen 
that the isolines of constant $q_\text{max}$ are roughly square-shaped which 
evidences that $q_\text{max}$ is a function of the largest time scale, i.e.\ 
$t_\text{max}\,{=}\,\max(t_e, t_d)$. This is confirmed in Fig.\,4(a) of the 
manuscript, where $q_\text{max}$ is plotted as a function of $t_\text{max}$ 
for different values of the structural parameters. The overall trend of 
$q_\text{max}$ can be roughly captured by a power-law scaling $q_\text{max}
\,{=}\,1{/}\sqrt{t_\text{max}}$. The observed deviations from the power-law 
scaling originate from the intrinsic stochasticities as well as our method of 
compression ratio calculation. For small values of $t_e$, $t_d$ and $p$ and 
large values of $n$, the points of the structural parameters domain are 
mapped onto a patch with low but nonzero extension along the $\Delta Q_r$ 
direction in the signal domain. By increasing the entering or emission 
time, the points collapse on a nonmonotonic curve which covers a much 
larger range along $\Delta Q_r$ direction; see Fig.\,S1(b). This results 
in larger $u_x$ or $u_y$ in the metric Eq.\,[\ref{Eq:ScaledMetric}], thus, 
smaller $C$ values. Hence, those points may be considered as non-invertible 
despite that their mapping to the signal domain is properly resolved. 
\smallskip\smallskip\smallskip

From the power-law relation between $q_\text{max}$ and $t_\text{max}$ 
and the fact that the time scales $t_e$ and $t_d$ are measured in 
units of the time resolution of observation $\Delta t$, it reads
\begin{equation}
q_\text{max} = \sqrt{\frac{\Delta t}{t_\text{max}}}.
\label{Eq:qmaxPowerLaw}
\end{equation}
According to Eq.\,[\ref{Eq:q1}], the value of $q$ for a given dendritic 
tree can be almost arbitrarily tuned through $\Delta t$. If the upper 
estimate of $q$ from Eq.\,[\ref{Eq:q1}] (obtained for a smooth channel, 
i.e.\ $V_\text{spines}{=}0$) is less than $q_\text{max}$ given by 
Eq.\,[\ref{Eq:qmaxPowerLaw}], variations of $q$ due to morphological 
changes of spines can be fully resolved with our proposed approach. 
By equating Eq.\,[\ref{Eq:qmaxPowerLaw}] with Eq.\,[\ref{Eq:q1}] at 
$V_\text{spines}{=}0$ we obtain the relation
\begin{equation}
t_\text{max} = \frac{1}{4\Delta t}\left(\frac{L^2}{D}\right)^2
\label{Eq:Tmax}
\end{equation}
between $t_\text{max}\,{=}\,\max(t_e,t_d)$, the time resolution $\Delta 
t$, and the diffusive timescale $\frac{L^2}{D}$ ($D$ is the diffusion 
coefficient of the tracer particles in the spineless dendritic channel). 
Therefore, for a given set of dendritic tree structure and tracer particle, 
the required time resolution of measurements $\Delta t$ is inversely proportional 
to $t_\text{max}$, i.e.\ the maximum time scale among the entering and emission 
times $t_e$ and $t_d$. In Fig.\,4(b) of the manuscript, $\Delta t$ is plotted 
versus the diffusive timescale $\frac{L^2}{D}$ for different values of 
$t_\text{max}$. The vertical lines mark the relevant range of $\frac{L^2}{D}$ 
for realistic values of the branching distance $L$ and diffusion coefficients 
$D$ for Ca\textsuperscript{2+}, fluorescein dextran (FD), and green fluorescent 
protein (GFP), as a few examples. It shows that a slower diffusion of tracer 
particles and shorter entering and emission times lead to a broader possible 
range for the time resolution of measurements $\Delta t$. For time resolutions 
around $\Delta t\,{\simeq}\,0.04\,\text{s}$ (typical for currently available 
cameras) and tracer particles with diffusion coefficients similar to GFP, the 
required entering and emission times can be up to a few minutes. However, in 
special techniques such as nuclear magnetic resonance spectroscopy, one deals 
with time resolutions of several seconds, demanding more slowly diffusing 
particles and shorter entering and emission times on a sub-minute scale.

\newpage

\begin{figure*}[b]
\begin{center}
\includegraphics[width=0.9\textwidth]{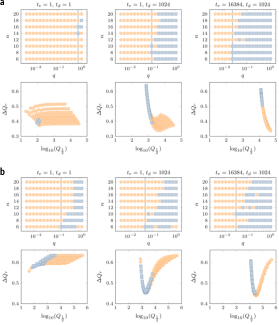} \\
\end{center}

\vspace{0.5em}

\parbox{\textwidth}{%
\justifying 
\noindent\textbf{Suppl.\,Fig.\,S1:}\ Visualisation of mapping the structural 
parameters $(q,n)$ to the signal characteristics $\big(\!\log\!_{_{10}}\!(
Q_{\frac12}\!), \Delta Q_r\!\big)$ for different entering times $t_e$ and 
emission times $t_d$. Other parameter values: (a) $p\,{=}\,0.55$, (b) $p\,
{=}\,0.45$. The points where the map is invertible (not invertible) for the 
compression threshold $C\,{=}\,0.05$ are shown with orange circles (blue 
squares). The blue vertical line marks $q_\text{max}$, up to which the map 
is invertible for all values of $n$.}
\label{Fig:S1}
\end{figure*}

\newpage

\begin{figure}[b]
\begin{center}
\includegraphics[width=0.5\textwidth]{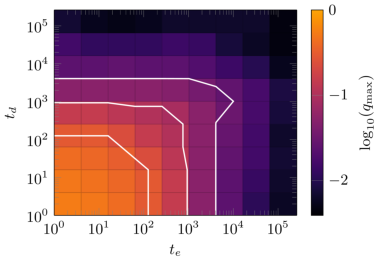} \\
\end{center}

\vspace{0.5em}

\parbox{\textwidth}{%
\justifying 
\noindent\textbf{Suppl.\,Fig.\,S2:}\ Logarithm of $q_\text{max}$ (i.e.\ the 
maximum value of $q$ up to which the map from $(q,n)$ to signal domain is 
invertible) as a function of entering and emission times $t_e$ and $t_d$. 
$q_\text{max}$ is extracted for $p\,{=}\,0.55$ and $C_\text{min}\,{=}\,
0.05$. The solid white lines represent isolines of constant $q_\text{max}$.}
\label{Fig:S2}
\end{figure}

\newpage

\begin{figure*}[b]
\begin{center}
\includegraphics[width=0.95\textwidth]{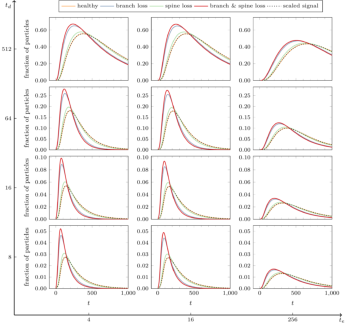} \\
\end{center}

\vspace{0.5em}

\parbox{\textwidth}{%
\justifying 
\noindent\textbf{Suppl.\,Fig.\,S3:}\ Fraction of particles in the soma as a 
function of time for healthy and differently degenerated dendritic trees for 
different entering and emission times $t_e$ and $t_d$. For the healthy dendrite 
(orange line) the following parameters were assumed: $n\,{=}\,10$, $\rho\,{=}
\,1\,\mu\text{m}^{-1}$, $V_\text{head}\,{+}\,V_\text{neck}\,{=}\,0.55\,\mu
\text{m}^3$ and $R\,{=}\,1\,\mu\text{m}$. The time resolution was chosen to 
be $\Delta t\,{=}\,\displaystyle\frac{L^2}{8D}$ corresponding to $\Delta t\,
{=}\,2.5\,\text{s}$ for a dendrite with mean branch length $L\,{=}\,20\,\mu
\text{m}$ and particles with a diffusion constant $D$ similar to GFP. The 
degeneracies were branch loss (blue line) where the tree has lost three 
generations of branches, spine loss (green line) where the dendrites have 
lost three quarters of their spine volume as well as the combination of 
both (red line). Increasing $t_d$ increases the fraction of particles in 
the soma leading to a broader and higher curve. Increasing $t_e$, on the 
other hand, leads to broader and flatter curves because of the restricted 
influx of particles. For the healthy dendrite, the signal $I{/}{\mathcal N}$ 
(generated by the accumulated pulses of the particles in the soma scaled by 
$t_d$ and shifted by one step to the left) is shown with the dotted line. 
This line coincides with the one for particle fraction, exemplifying that 
the particle fraction in the soma can be obtained by the signal and vice 
versa.}
\label{Fig:S3}
\end{figure*}

\end{widetext}

\end{document}